\documentstyle[aps,prl,multicol,epsf]{revtex}

\def\date#1{\author{\small(#1)}}
\def\abstract#1{\author{\parbox[t]{5.5in}{\small#1}}\par\maketitle}

\def\f.#1.{{\bf #1}}
\def\mb.#1.{\bbox{#1}}

\def\opn{\begin{equation}} \def\cls{\end{equation}}
\def\opa{\begin{eqnarray}} \def\cla{\end{eqnarray}}
\def\opbib{\begin{references}} \def\clbib{\end{references}}
\def\bb#1{\bibitem{#1}}
\def\qno;#1;{\label{#1}\end{equation}} \def\qna;#1;{\label{#1}\end{eqnarray}}
\def\rf;#1;{(\ref{#1})}
\def\dels#1 {\nabla\kern -1.5pt_{#1}\kern 1.5pt}
 \def\av#1{\left\langle #1\right\rangle}

\def\suspend{\end{multicols}\vspace*{-0.5cm} \noindent \rule{8.65cm}{.02cm}}
\def\resume{\hskip 9.3cm \rule{8.65cm}{.02cm} 
\begin{multicols}{2}\vspace*{-0.8cm} \noindent}
\def\umo{\rlap{\"{$\,$\ }} \,\,\,\kern -.3 cm o}

\newcount\hours \newcount\minutes \newcount\a \newcount\b
\def\gmt{\hours = \time \divide\hours by 60 \a =\hours \multiply \a by 60
\minutes = \time \advance \minutes by -\a
{\ifnum\hours<10 0\fi}\number\hours
{\ifnum\minutes<10 0\fi}\number\minutes}
\def\gday{\b=\year \advance \b by -1900
\number\b{\ifnum\month<10 0\fi}\number\month{\ifnum\day<10 0\fi}\number\day}
\def\today{\number\day\ 
\ifcase\month\or January\or February\or March\or April\or May\or 
June\or July\or August\or September\or October\or November\or December\fi
\ \number\year}

 \def\be{\beta}  \def\de{\delta}
 \def\ze{\zeta} \def\et{\eta} \def\th{\theta} 
\def\ka{\kappa}  \def\rh{\rho}  
 \def\ph{\phi}

\def\casefr#1/#2 {\case{#1}{#2}}
\def\part{\partial} \def\del{\nabla}

\def\half{\case{1}{2}}  

 \def\bdel{\mb.\del.} \def\prop{\propto} 
   
\def\bx{\f.x.}   \def\bu{\f.u.} 
 \def\br{\f.r.}  
   
\def\bcdot{\mb.\cdot.} 
\def\trademark{$^{\rlap{$\scriptstyle\bigcirc$}\,\mbox{\tiny R}}$\ }
\def\sump#1{\lower .15in\hbox{$\stackrel{\displaystyle{\sum}'} {\scriptstyle 
#1}$}}

\begin{document}

\title{\vbox to 0pt {\vskip -1cm \rlap{\hbox to \textwidth {\rm{\small PHYS. 
REV. LETT., 
\f.78. 4922-4925 (1997)} \hfill chao-dyn/9611007}}}Passive Scalar: Scaling 
Exponents 
and Realizability}

\author{Robert H. Kraichnan$^*$}

\address{369 Montezuma 108, Santa Fe, NM 87501-2626}

\date{Revised 7 April 1997}

\abstract{An isotropic passive scalar field $T$ advected by a rapidly-varying 
velocity field is studied. The tail of the probability distribution $P(\th,r)$ 
for the difference $\th$ in $T$ across an inertial-range distance $r$ is found 
to be Gaussian. Scaling exponents of moments of $\th$ increase as $\sqrt{n}$ 
or faster at large order $n$, if a mean dissipation conditioned on $\th$ is a 
nondecreasing function of $|\th|$. The $P(\th,r)$ computed numerically under 
the so-called linear ansatz is found to be realizable. Some classes of gentle 
modifications of the linear ansatz are not realizable.}

\vskip .4cm

\begin{multicols}{2}

In the past several years, a number of papers have dealt with the theory of a 
statistically isotropic passive scalar field advected by an incompressible 
velocity field that varies very rapidly in time (white velocity field) 
\cite{1,2,3,4,5,6,7,8,9,10,11,12,13,14,15,16,17,18,19,20,21,22,23,24,25,26}. 
Some exact statistical relations can be written for this problem, which leads 
to the hope that, for the first time, anomalous scaling exponents for the 
inertial range can be derived analytically in a turbulence problem. 

Nevertheless, progress has been limited. Closed equations exist for each 
order of $n$-point moments of the scalar field \cite{3}, but above 2nd order 
the number of independent variables is daunting. Perturbation analysis has 
yielded predictions of finite-order scaling exponents at two remote borders of 
the problem: infinite space dimensionality \cite{6,7}; and infinitely nonlocal 
interaction of spatial scales of the velocity and scalar fields, with 
diffusivity exponent $\xi\to0$ in (10) below \cite{9,14}. Neither regime is 
accessible to test by direct numerical simulation (dns).

A physically-motivated ``linear ansatz'' for dissipation terms in the 
equation of motion \cite{1,10} predicts all exponents for the full domain of  
$d$ and $\xi$. It has quantitative support from dns \cite{10,27} and from some 
experiments \cite{20}. The ansatz has not been derived analytically. The 
predictions in the limit $\xi\to0$ are implausible and in conflict with the 
perturbation results. It is unclear whether the ansatz has any domain of exact 
validity.

In the present paper it is deduced that, contrary to common expectation, the 
tail of the probability distribution function (pdf) of spatial scalar-field 
differences is Gaussian in the inertial range, rather than exponential or 
stretched-exponential. This leads to a relation between the asymptotic 
behavior of scaling exponents and that of a conditional mean of dissipation. 
The linear ansatz is found to yield a realizable (everywhere positive) pdf, 
while some gentle deviations from it do not. A general apparatus is formulated 
for relating scaling exponents to the pdf shape in the inertial range. It is 
used to obtain the linear-ansatz pdf explicitly.

The passive scalar field $T(\bx,t)$ obeys
\opn
\left( {\part\over\part t} + \bu(\bx,t)\bcdot\bdel \right)T(\bx,t) = 
\ka\del^2 T(\bx,t).
\qno;1;
where  $\ka$ is molecular diffusivity. A statistically steady state may be 
maintained by adding an appropriate source term to \rf;1;

The structure functions are defined by $S_q(r) = \av{|\th(\br)|^q}$, where 
$\th(\br)$ denotes $T(\bx+\br,t) - T(\bx,t)$ and $\av{ }$ denotes ensemble 
average over homogeneous, isotropic statistics. Suppose that there is 
power-law inertial-range scaling of the structure functions $S_q(r)$,
\opn
S_q(r) \prop (r/L)^{\ze_q},
\qno;2;
where $L \gg r \gg \ell_d$, $L$ is a large length scale, the order of initial 
or injection scales, and $\ell_d$ is a dissipation scale. A realizable 
(everywhere positive) pdf $P(\th,r)$ for $\th(\br)$ requires that $d\ze_q/dq$ 
be a non-increasing function of $q$ (H\"older inequalities). This does not 
exhaust the realizability conditions on $\ze_q$. Necessary and sufficient 
conditions for an everywhere non-negative $P(\th,r)$, nonzero at an infinite 
set of $\th$ values, are \cite{28}
\opn
\det[S_{i+j}(r)]_{i,j=0,1,...,n} > 0, \quad (n=0,1,2,...)
\qno;3;
where $S_{i+j}$ is set to zero for odd $i+j$. $P(\th,r)$ is unique if, in 
addition, $\sum_0^\infty [S_{2n}(r)]^{-1/2n} = \infty$ (Carleman's criterion 
\cite{28}). This is satisfied if $P(\th,r)$ falls off exponentially or faster 
as $|\th|\to\infty$.

Direct calculation of the moments of $P(\th,r)$ verifies that the $r$ 
dependence implied by \rf;2; is
\opn
P(\th,r) = \int_0^\infty \rh\left(a,{r\over r'}\right)P\left({\th\over 
a},r'\right){da\over a},
\qno;4;
where $r$ and $r'$ are both in the scaling range and $\rh(a,R)$ has moments 
$\int_0^\infty a^{2n}\rh(a,R)da = R^{\ze_{2n}}$. Explicitly,
\opn
\rh(a,R) = {2\over\pi} \int_0^\infty \cos(ka)\ph(k,R)dk,
\qno;5;
\opn
\ph(k,R) = 1 + \sum_{n=1}^\infty R^{\ze_{2n}}(-k^2)^n/(2n)!.
\qno;6;
Eqs. \rf;4;--\rf;6; are easily generalized to the case where the 
$S_{2n}(r)/S_{2n}(r')$ do not scale as powers of $r/r'$.

The H\"older relations are equalities for normal scaling $\ze_{2n} = n\ze_2$, 
which yields $\rh(a,R) = \de(a-R^{\ze_2/2})$. The smallest possible values of 
the higher $\ze_{2n}$ plausibly are $\ze_{2n} = \ze_2$, which yields $\rh(a,R) 
= (1-R^{\ze_2})\de(a-(+0)) + R^{\ze_2}\de(a-1)$. This corresponds to regions 
of constant $T$ interspersed with regions where $\th(\br)$ is as large as the 
macroscale differences $\th(L)$; \rf;1; cannot magnify differences in the 
injected $T$, only change their spatial scale and relax them. Unless the 
scaling is normal, $\rh(a,R)$ increases in normalized width $a/R^{\ze_2/2}$ as 
$R$ decreases, so that $P(\th,r)$ cannot have a similarity form like 
$r^{-\be}F(\th/r^\be)$.

In both cases $\ze_{2n} = n\ze_2$ and $\ze_{2n}=\ze_2$, the tail of 
$P(\th,r)$ is Gaussian if $P(\th,r)$ is Gaussian for $r \sim L$. In a Gaussian 
pdf $\ln S_{2n} \sim n\ln n$ for large $n$, to within terms of order $n$. If 
$\ln S_{2n}(r) \sim n\ln n$ for $r \sim L$, it follows that, for all $n\ze_2 
\ge \ze_{2n} \ge \ze_2$, $\ln S_{2n}(r) \sim n\ln n$ for large $n$ throughout 
the scaling range. This is asymptotically Gaussian behavior of the tail. An 
exponential or stretched exponential pdf with tail $\ln P(\th,r) \propto 
-|\th|^b$, with $b < 2$, instead requires $\ln S_{2n} \sim (2n/b)\ln n$ as 
$n\to\infty$. Thus the tail of $P(\th,r)$ cannot fall off more slowly than 
Gaussian in the scaling range. This does not preclude exponential or slower 
fall-off in the dissipation range of $r$. The same result holds for the pdf of 
velocity differences in the inertial range of Navier-Stokes turbulence. Other 
arguments for faster-than exponential tails in the inertial range have 
recently been given by Ching and Procaccia and by Noulez \cite{29}.

In interpreting this result, a distinction must be made between the presence 
of a central cusp in the pdf, expressing the existence of regions in which 
there is almost no excitation, and the shape of the far tail. Confusion 
between these two features can result in spurious fitting of a pdf to an 
exponential or stretched exponential. An example of a cusped pdf is that 
implied by the linear ansatz and shown in Fig. 2.

If there is a range of $r$ where the source term may be neglected, the 
steady-state balance equation for the even-integer structure functions 
$S_{2n}(r)$ is \cite{1}
\opn
- {2\over r^{d-1}}{\part\over\part r} \left( r^{d-1} \et(r) {\part 
S_{2n}(r)\over\part r} \right) = \ka J_{2n}(r),
\qno;7;
in the limit of rapidly-changing (white) velocity field. The left side of 
\rf;7; is derived perturbatively from the $\bu\cdot\bdel$ term in \rf;1; in 
the white limit. Here $d$ is space dimensionality, $\eta(r)$ is the 
two-particle eddy-diffusivity scalar exerted by the velocity field, and
\opn
J_{2n}(r) = 2n\av{[\th(\br)]^{2n-1} H[\th(\br)]},
\qno;8;
\opn
H[\th(\br)] = \av{(\del_x^2 + \del_{x'}^2)\th(\br)|\th(\br)},
\qno;9;
where $\av{\bcdot|\th(\br)}$ denotes ensemble average conditioned on a given 
value $\th(\br)$, and $\bx'=\bx+\br$.

Power-law dependence of $S_2(r)$ in the inertial range is assured from the 
exact, closed equation of motion for $S_2(r)$, provided that $\et(r)$ has the 
form
\opn
\et(r) = \et(L)(r/L)^{\xi} \quad (0 < \xi < 2),
\qno;10;
for $L \gg r \gg \ell_d$. The steady-state scaling exponent for $S_2(r)$ 
found from \rf;7; is $\ze_2 = 2 - \xi$.

There is no a priori assurance of power-law inertial-range scaling of 
$S_{2n}(r)$ for $n>1$. If it exists, the balance of advection and $\ka$ terms 
in \rf;7; implies that the $J_{2n}(r)$ have the form
\opn
J_{2n}(r) = nC_{2n}J_2(r)S_{2n}(r)/S_2(r),
\qno;11;
where the $C_{2n}$ are dimensionless constants and
\opn
C_{2n} ={\ze_{2n}(\ze_{2n} + d - \ze_2)\over nd\ze_2}.
\qno;12;
The H\"older inequalities $\ze_{2n} \le n\ze_2$ give
\opn
C_{2n} \le 1 + {n-1\over d}\ze_2.
\qno;13;
Equality in \rf;13; correponds to normal scaling $\ze_{2n} = n\ze_2$.

The simplest candidate for an anomalous scaling solution to \rf;7; is the 
``linear ansatz'' \cite{1}:
\opn
H(\th,r) = J_2(r)\th/S_2(r).
\qno;14;
It corresponds to $C_{2n}=1$ for all $n$ and yields
\opn
\ze_{2n} = \half\sqrt{4nd\ze_2 + (d-\ze_2)^2} - \half(d-\ze_2).
\qno;15;
 Linear relations like \rf;14; were earlier proposed by Ching and Pope 
\cite{30,31}.

Using Mathematica\trademark 3.0, I have verified numerically, for a number of 
values of $d$ and $\ze_2$, that \rf;15; satisfies \rf;3; up to $2n=100$, even 
in the extreme case where $P(\th,r')$ in \rf;4; is taken as a $\de$ function. 
The $\ze_{2n}$ corresponding to $C_{2n} = 1 + \be(n-1)$ appear to satisfy 
\rf;3; for $0 < \be\le\ze_2/d$. This case \cite{23} yields $\ze_{2n} \prop n$ 
as $n\to\infty$; $\be=\ze_2/d$ is normal scaling. The perturbation analyses 
\cite{6,7,9,14} give $n\ze_2-\ze_{2n}$ that are quadratic in $n$ for moderate 
$n$. They also satisfy \rf;3;; the full set of $\ze_{2n}$, and hence 
$P(\th,r)$, are not predicted.

A conditional mean related to dissipation may be defined by
\opn
K(\th,r) = \av{|\bdel_x\th + \bdel_{x'}\th|^2|\th}.
\qno;16;
If $r \gg \ell_d$, the $\bdel_x\bdel_{x'}$ crossterm in \rf;16; is 
negligible, and $K(\th,r) \approx \av{|\bdel_x\th|^2 + 
|\bdel_{x'}\th|^2|\th}$, the conditional mean of the average of the 
dissipation at the points $\bx$ and $\bx'$. $H(\th,r)$, $K(\th,r)$, and 
$P(\th,r)$ are related by an identity that follows from homogeneity alone 
\cite{31,32}:
\opn
H(\th,r)P(\th,r) \equiv {\part\over\part\th}[K(\th,r)P(\th,r)].
\qno;17;
If the tail of $P(\th,r)$ is Gaussian, and $K(\th,r)$ is no stronger than a 
power of $|\th|$, the leading term on the right of \rf;17; at large $|\th|$ is 
$\prop -\th K(\th,r)P(\th,r)$, whence $H(\th,r)/\th \prop -K(\th,r)$.

Monotonic growth of $K(\th,r)$ with $|\th|$ at inertial-range $r$ is a very 
weak, qualitative form of Kolmogorov refined similarity hypothesis 
\cite{33,34}. If it holds, then the linear ansatz for $H(\th,r)$ and 
associated $\ze_{2n} \prop n^{1/2}$ behavior at large $n$ represent a lower 
bound for the asymptotic growth of $\ze_{2n}$. This bound is achieved if 
$K(\th,r)$ tends to a constant as $|\th|\to\infty$. Sublinear growth of 
$H(\th,r)$ at large $|\th|$ would make $K(\th,r)$ a decreasing function of 
$|\th|$ and would yield slower growth of $\ze_{2n}$ than $n^{1/2}$ at large 
$n$. Yakhot \cite{13} and Chertkov \cite{17} predict $\ze_{2n} \to$ constant 
as $n\to\infty$, which can be shown to imply $K(\th,r) \prop |\th|^{-2}$ as 
$|\th|\to\infty$ [see \rf;19; and \rf;22;].

The analytical relations between $H(\th,r)$, $P(\th,r)$, and the $\ze_{2n}$ 
in the inertial range may be expressed in terms of a ``co-pdf'' $P_H(\th,r)$ 
with moments $\int_{-\infty}^{\infty}\th^{2n}P_H(\th,r)d\th = C_{2n}S_{2n}(r)$ 
$(C_0 \equiv 1)$. Then by \rf;8; and \rf;9;,
\opn
{H(\th,r)\over\th}={J_2(r)\over S_2(r)}{P_H(\th,r)\over P(\th,r)},
\qno;18;
and $\int_{-\infty}^\infty P_H(\th,r)d\th = 1$. If \rf;3; is satisfied with 
$S_{2n}(r)$ replaced by $S^H_{2n}(r)=C_{2n}S_{2n}(r)$, then $P_H$ and 
$H(\th,r)/\th$ are everywhere positive. By \rf;17;,
\opn
P_H(\th,r) \equiv {1\over\th}{S_2(r)\over J_2(r)}{\part\over\part\th} 
[K(\th,r)P(\th,r)].
\qno;19;

$P_H(\th,r)$ is related to functions $\rh_H(a,r)$ and $\ph_H(k,r)$ by 
equations identical to \rf;4;--\rf;6; except that a factor $C_{2n}$ appears in 
$\ph_H(k,r)$:
\opn
\ph_H(k,R) = 1 + \sum_{n=1}^\infty R^{\ze_{2n}}C_{2n}(-k^2)^n/(2n)!.
\qno;20;

$P_H(\th,r) = P(\th,r)$ under the linear ansatz. $P_H(\th,r)$ can be found 
analytically in some other cases that are formal solutions of \rf;12;. One is 
$C_{2n} = 1 + \be(n-1)$. The associated $P_H$ is
\opn
P_H(\th,r) = \left( 1 - {3\be\over2} \right) P(\th,r) -{\be\th\over2} {\part 
P(\th,r)\over \part\th}.
\qno;21;
Another case is $\ze_{2n}=\ze_2$ at all $n$, for which \rf;12; yields 
$C_{2n}=1/n$. These $C_{2n}$ values are generated by
\opn
P_H(\th,r) = {1\over|\th|}\int_{|\th|}^\infty P(\th',r)d\th'.
\qno;22;
More generally, closed forms for $P_H$ can be written if $C_{2n}$ is a finite 
polynomial in $n$.

Eqs. \rf;6; and \rf;20; converge rapidly because of the denominators $(2n)!$. 
They have infinite radii of convergence if the $\ze_{2n}$ satisfy the H\"older 
inequalities. In some cases it is feasible to sum \rf;6; and \rf;20; 
numerically, perform the transforms and explore the shapes of $P$ and $P_H$. 
Both $\ph(k,R)$ and $\ph_H(k,R)$ can be computed for large values of $k$, far 
out in the tails, by the use of Mathematica\trademark to sum hundreds of terms 
at precisions of the order of a hundred decimal digits.

\hspace{.3cm}
\epsffile{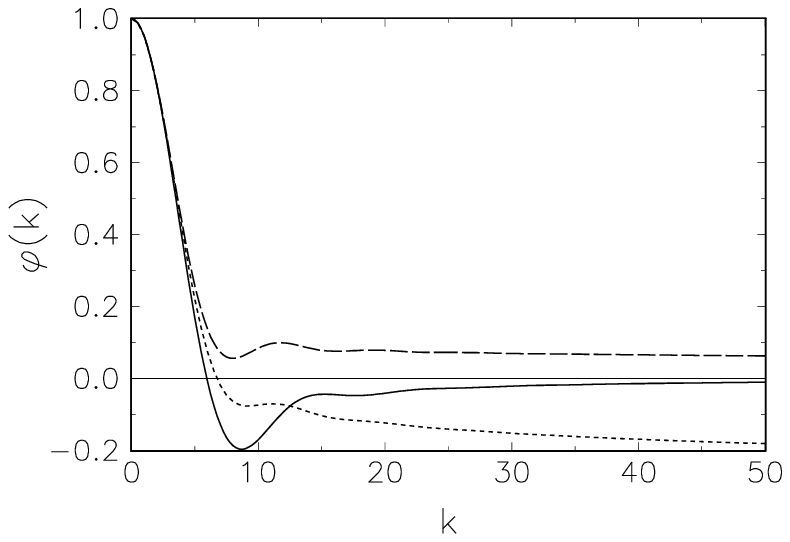}
\vspace{.8cm}

\small\baselineskip.12in
\noindent FIG. 1. $\ph(k,R)$ for $d=2$, $\ze_2=1/2$, $R=0.01$ under the 
linear ansatz (solid line). The dashed line is $\ph(k,R)$ if the 
linear-ansatz 
$\ze_{2n}$ are reduced by $n^{1/10}$, and the dotted line is $\ph_H(k,R)$ for 
that case.

\normalsize\baselineskip.15in
\vspace{.5cm}

\hspace{.3cm}
\epsffile{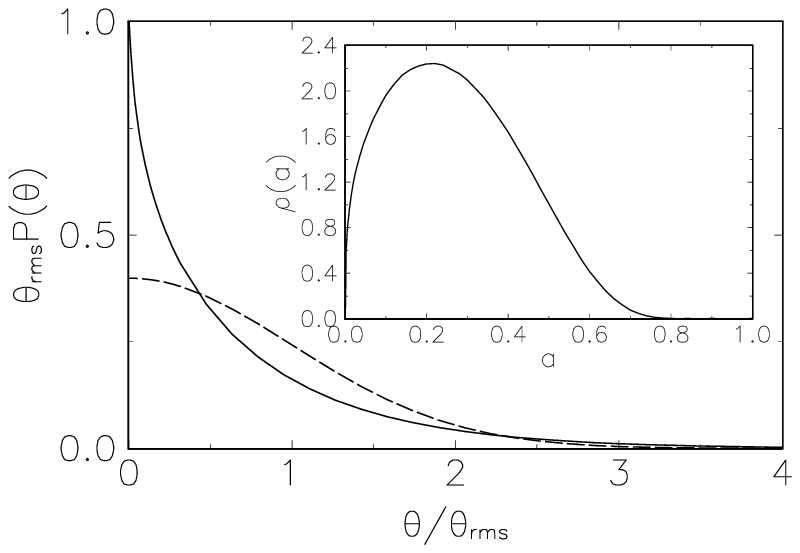}
\vspace{.8cm}

\small\baselineskip.12in
\noindent FIG. 2. $P(\th,r)$ for the linear-ansatz case (solid line); 
$\th_{rms}=10^{-1/2}$. The dashed curve is a normalized Gaussian. The inset 
is 
$\rh(a,R)$ for the linear-ansatz case.

\normalsize\baselineskip.15in
\vspace{.5cm}

Fig. 1 shows $\ph(k,R)$ with $R=1/100$ and $\ze_{2n}$ given by the 
linear-ansatz at $d=2$, $\ze_2=1/2$. Fig. 2 shows $\rh(a,R)$ and $P(\th,r)$ as 
constructed from directly calculated $\ph(k,R)$ values for $0 \le k \le 200$ 
and an extrapolation for $k > 200$. The local exponent at $k=200$ is $-1.367$ 
and is slowly decreasing. $P(\th,r')$ is taken as a Gaussian.

One can argue that, if the linear ansatz is self-consistent, the branch point 
in \rf;15; at
\opn
\ze_{q_c} = -(d-\ze_2)/2, \quad q_c = -(d-\ze_2)^2/(2d\ze_2).
\qno;23;
should mark the limit of applicability of \rf;15;, and $q_c$ should represent 
the most-negative moment order that exists \cite{35}. $P(|\th|,r')$ is here 
taken as a $\de$ function so that its moments of all negative orders exist. I 
therefore conjecture that the ansatz gives $\ph(k,R) \prop k^{-|q_c|}$ 
$(k\to\infty)$ so that $\rh(a,R) \prop a^{|q_c|-1}$ at the origin. The 
$P(\th,r)$ given by \rf;4; then is $\prop$ const.$-|\th|^{|q_c|-1}$ at the 
origin if $3 > |q_c| > 1$ and $\prop |\th|^{|q_c|-1}$ (infinite cusp) if 
$|q_c| < 1$. The crossover is at $\ze_2^C = (2-\sqrt{3})d$. The data shown in 
Fig. 2 appear to be consistent with exponent $-1.125$ at $k=\infty$, together 
with $\int_0^\infty \ph(k,R) dk = 0$, in accord with this conjecture.

The primary fact here is that the linear ansatz appears to yield a $\rh(a,R)$ 
that is positive everywhere, ensuring positive $P(\th,r)$ also. The positivity 
of $\rh(a,R)$ is a stronger result than \rf;3;. It implies positive $P(\th,r)$ 
even in the unphysical case where $P(\th,r')$ is a $\de$ function. In Fig. 2 
there is an absolute cutoff of $\rh(a)$ at finite $a$. This confirms that 
$P(\th,r)$ is Gaussian-like at large $|\th|$, despite its appearance, if 
$P(\th,L)$ is Gaussian.

Fig. 1, dashed curve is the $\ph(k,R)$ that results from a gentle 
modification of these linear-ansatz exponents, $\ze_{2n} = 
\ze_{2n}^Ln^{-1/10}$, where $\ze_{2n}^L$ satisfies \rf;15;. This leaves 
$\ze_2$ unchanged and makes the $\ze_{2n}$ go as $n^{2/5}$ as $n\to\infty$. 
The large-$k$ ($k \sim 200$) tail of $\ph$ appears to go approximately as 
$k^{-0.18}$; \rf;3; is satisfied for $2n \le 100$. The implied behavior of $P$ 
at small $\th$ is $\th^{-0.18}$. The result for $\ph_H(k,r)$ is Fig. 1, dotted 
curve. Its large-$k$ behavior is qualitatively different from that of 
$\ph(k,R)$; $\ph_H$ continues to grow more negative at $k=200$. Its tail is 
very different from that of $\ph$, which suggests bizarre behavior of 
$H(\th,r)$ at small $\th$. The effective exponents $\ze_{2n}^H = \ze_{2n} + 
C_{2n}\ln R$ for this case differ little from the $\ze_{2n}$. If they are 
tried as exponents $\ze_{2n}$, they satisfy the H\"older inequalities, but 
they violate \rf;3; at $2n=56$ if $P(\th,r')$ in \rf;4; is taken as a $\de$ 
function. This illustrates that mild and innocent-looking deviations from the 
linear ansatz can destroy realizability.

Numerical investigation is feasible for other values of $\ze_2$ and $R$, and 
other deviations from the linear ansatz (e.g. $\ze_{2n}$ behaving as $n^{1/2}$ 
at large $n$, but with $C_{2n} \not= 1$). The results appear to be consistent 
with the conjecture that \rf;23; determines the behavior of $P(\th,r)$ at 
$\th=0$ under \rf;15;. In particular, for $d=2$, $\ze_2 = 1$, $R=0.01$, the 
large-$k$ exponent of $\ph(k,R)$ clearly lies between $0$ and $-1$ 
($\ze_q^C=0.536$). Sensitivity to changes of the $C_{2n}$ away from unity 
seems to be widespread.

I am grateful to Shiyi Chen, E. Ching, G. Eyink, T. Gotoh, I. Procaccia, K. 
Sreenivasan, and V. Yakhot for helpful interactions. This work was supported 
by The Department of Energy.

\vskip -.25in

\opbib

\vbox to -.6in { }

\bibitem[*]{0} Internet address: rhk@lanl.gov

\bb{1} R. H. Kraichnan, Phys. Rev. Lett. \f.72., 1016 (1994).

\bb{2} V. S. L'vov, I. Procaccia, and A. L. Fairhall, Phys. Rev E, \f.50., 
4684 (1994).

\bb{3} B. Shraiman and E. D. Siggia, Phys. Rev. E, \f.49., 2912 (1994).

\bb{4} E. Balkovsky, M. Chertkov, I. Kolokolov, and V. Lebedev, JETP Lett. 
\f.61., 1012 (1995).

\bb{5} M. Chertkov, G. Falkovich, I. Kolokolov, and V. Lebedev, Phys. Rev. E, 
\f.51., 5609 (1995).

\bb{6} M. Chertkov, G. Falkovich, I. Kolokolov, and V. Lebedev, Phys. Rev. E, 
\f.52., 4924 (1995).

\bb{7} M. Chertkov and G. Falkovich, Phys. Rev. Lett. \f.76., 2706 (1996).

\bb{8} A. Fairhall, O. Gat, V. L'vov, and I. Procaccia, Phys. Rev E, \f.53., 
3518 (1996).

\bb{9} K. Gaw\c edzki and A. Kupiainen, Phys. Rev. Lett. \f.75., 3834 (1995).

\bb{10} R. H. Kraichnan, V. Yakhot, and S. Chen, Phys. Rev. Lett. \f.75., 240 
(1995).

\bb{11} V. L'vov and I. Procaccia, Phys. Rev. Lett. \f.76., 2898 (1996).

\bb{12} A. Pumir, preprint 1995.

\bb{13} V. Yakhot, submitted to Phys. Rev. E, 1995 and 1996.

\bb{14} D. Bernard, K. Gaw\c edzki, and A. Kupiainen, preprint 
chao-dyn/9601018.

\bb{15} M. Chertkov, G. Falkovich, and V. Lebedev, Phys. Rev. Lett. \f.76., 
3707 (1996).

\bb{16} D. Segel, V. L'vov, and I. Procaccia, Phys. Rev. Lett. \f.76., 1828 
(1996).

\bb{17} M. Chertkov, submitted to Phys. Rev. E (chao-dyn/9606011).

\bb{18} E. Balkovsky, G. Falkovich, I Kolokolov, and V. Lebedev, submitted to 
Phys. Rev. E (chao-dyn/9603015).

\bb{19} E. S. C. Ching, Phys. Rev. E \f.53., 5899 (1996)

\bb{20} E. S. C. Ching, V. S. L'vov, Evgeni Podivilov, and I. Procaccia, 
Phys. Rev. E \f.54., 6364 (1996) (chao-dyn/9608008).

\bb{21} O. Gat, V. S. L'vov, E. Podivilov, and I. Procaccia, submitted to 
Phys. Rev. Lett.

\bb{22} G. L. Eyink, Phys. Rev. E \f.54., 1497 (1996).

\bb{23} E. S. C. Ching, V. S. L'vov, and I. Procaccia, Phys. Rev. E \f.54., 
R4520 (1996) (chao-dyn/9606017).

\bb{24} G. Eyink and J. Xin, Phys. Rev. Lett. \f.77., 2674 (1996).

\bb{25} R. Benzi, L. Biferale, and A. Wirth (chao-dyn/9702005).

\bb{26} M. Vergassola and A. Mazzino (chao-dyn/9702014).

\bb{27} S. Chen and R. H. Kraichnan, in preparation

\bb{28} H. S. Wall, {\it Analytic Theory of Continued Fractions} (New York, 
Van Nostrand, 1948), Secs. 86-88. Example: the modified Gaussian moments 0.9, 
3, 15, 105, 945, ... satisfy all H\"older inequalities but violate \rf;3; at 
$2n=22$.

\bb{29} A. Noulez, unpublished; E. S. C. Ching and I. Procaccia, unpublished.

\bb{30} E. S. C. Ching, Phys. Rev. Lett. \f.70., 283 (1993).

\bb{31} S. B. Pope and E. S. C. Ching, Phys. Fluids A, \f.5., 1529 (1993).

\bb{32} E. S. C. Ching and R. H. Kraichnan, in preparation

\bb{33} G. Stolovitzky and K. R. Sreenivasan, Rev. Mod. Phys. \f.66., 229 
(1994); G. Stolovitzky, P. Kailasnath, and K. R. Sreeivasan, J. Fluid Mech. 
\f.297., 275 (1995).

\bb{34} Y. Zhu, R. A. Antonia, and I. Hosokawa, Phys. Fluids \f.7., 1637 
(1995).

\bb{35} V. Yakhot, private communication.

\clbib

\end{multicols}

\end{document}